\documentclass[5p,twocolumn]{elsarticle}

\usepackage{amsmath}
\usepackage{amssymb}
\usepackage{acronym}
\usepackage{xcolor}
\usepackage{caption}
\usepackage{subcaption}
\usepackage[frak=esstix]{mathalpha}
\usepackage{algorithm}
\usepackage{algpseudocode}

\begin{document}

\acrodef{PB}{protective behavior}
\acrodef{GR}{government regulation}
\acrodef{NPI}{non-pharmaceutical intervention}
\acrodef{VPB}{voluntary protection behavior}
\acrodef{PB}{protection behavior}
\acrodef{SIR}{susceptible-infectious-removed}
\acrodef{ODE}{ordinary differential equation}
\acrodef{HBM}{Health Belief Model}
\acrodef{WHO}{World Health Organization}
\acrodef{OWID}{Our World in Data}
\acrodef{VoC}{variants of concern}
\acrodef{FIR}{finite impulse response}

\newcommand{\red}[1]{{\color{red}{#1}}}
\newcommand{\bh}[1]{{\color{blue}{\textbf{BH}: #1}}}

\newcommand{\deriv}[2]{\frac{\mathrm{d}#1}{\mathrm{d}#2}}

\newcommand{\cmax}{c^{\max}}
\newcommand{\imax}{i^{\max}}
\newcommand{\imaxsir}{i^{\max}_{\mathrm{SIR}}}
\newcommand{\ptwomax}{p_2^{\max}}

\newcommand{\Rt}{R(t)}
\newcommand{\istar}{i_*}
\newcommand{\pstar}{p_*}
\newcommand{\tstar}{t_*}
\newcommand{\tistar}{t_{i,*}}
\newcommand{\tpstar}{t_{p,*}}
\newcommand{\ystar}{y_*}

\newcommand{\iinfty}{i_{\infty}}
\newcommand{\pinfty}{p_{\infty}}

\newcommand{\niterations}{n_{\mathrm{iterations}}}
\newcommand{\stepsize}{\Delta_{\mathrm{x}}}

\newcommand{\rsir}{r^{\mathrm{SIR}}}
\newcommand{\ssir}{s^{\mathrm{SIR}}}

\newcommand{\tobs}{t_{\mathrm{obs}}}
\newcommand{\tsim}{t_{\mathrm{sim}}}
\newcommand{\ReLU}[1]{ \mathrm{ReLU}\left( #1 \right)}

\newcommand{\mchangeobs}{\Delta_{\mathrm{m}}^{\mathrm{obs}} }
\newcommand{\mobs}{m^{\mathrm{obs}} }
\newcommand{\mobsbaseline}{m^{\mathrm{base}} }

\newcommand{\logten}[1]{\log_{\mathrm{10}}\left( #1 \right)}
\newcommand{\dt}{\mathrm{d}t}

\newcommand{\transpose}{\mathrm{T}}
\newcommand{\jacobian}{\mathcal{J}}
\newcommand{\x}{\mathbf{x}}

\begin{frontmatter}

\title{Unraveling the role of adapting risk perception during the COVID-19 pandemic in Europe}

\author[l1]{Bastian Heinlein}
\ead{bastian.heinlein@fau.de}
\address[l1]{ {Friedrich-Alexander-Universität Erlangen-Nürnberg}, {Germany}}
\cortext[cor1]{Corresponding author}

\author[l2,l3,l4]{Manlio De Domenico\corref{cor1}}
\ead{manlio.dedomenico@unipd.it}
\address[l2]{ {Department of Physics and Astronomy "Galileo Galilei", University of Padua}, {Via F. Marzolo 8}, {315126 Padova}, {Italy} }
\address[l3]{ {Padua Center for Network Medicine, University of Padua}, {Via F. Marzolo 8}, {315126 Padova}, {Italy} }
\address[l4]{ {Istituto Nazionale di Fisica Nucleare, Sez. Padova}, {Italy} }

\begin{abstract}
During the COVID-19 pandemic, the behavioral response to reported case numbers changed drastically over time. While a few dozen cases were enough to trigger government-induced and voluntary contact reduction in early 2020, less than a year later, much higher case numbers were required to induce behavioral change. Little attention has been paid to understand, and mathematically model, this effect of decreasing risk perception over longer time-scales. Here, first we show that weighing the number of cases with a time-varying factor of the form $t^{a}\;,\;a<0$ explains real-world mobility patterns from several European countries during 2020 when introduced into a very simple behavior model. Subsequently, we couple our behavior model with an SIR epidemic model. Remarkably, decreasing risk perception can produce complex dynamics, including multiple waves of infection. We find two regimes for the total number of infected individuals that are explained by the interplay of initial attention and the rate of attention decrease. Our results show that including adaption into non-equilibrium models is necessary to understand behavior change over long time scales and the emergence of non-trivial infection dynamics. 
\end{abstract}


\end{frontmatter}

\section{Introduction}
\label{sec:introduction}
The behavioral response to the COVID-19 pandemic changed considerably over time. While relatively few reported cases were enough to trigger massive mobility reductions in March 2020, a much higher reported prevalence was necessary to elicit a similar response at the end of the year. Now, adherence to recommended \acp{NPI} dropped dramatically since the \ac{WHO} declared that COVID-19 no longer poses an health emergency of international concern \cite{who_end_health_emergency}. Yet, only a few attempts have been made to explain these drastic changes of the behavioral response.  

One reason reason for this might be the lack of longitudinal data about behavior change. Even though some studies collected survey data over longer periods \cite{facebook_survey,betsch_COSMO,sune_lehmann_comm_med,wambua_CoMix_study}, they miss relevant phases of the pandemic, especially the period when the first cases were reported. Furthermore, the used questionaires typically changed over time, making analyses more difficult.

To overcome this limitation we propose to interpret mobility data as a proxy for behavior change. In contrast to surveys, mobility data is continuously captured from smartphones and wearable devices in most countries. Aggregate data about mobility reduction during the COVID-19 pandemic has been readily provided, e.g., by Google \cite{google} and Apple \cite{apple}. Conveniently, these data are not subject to biases typical of self-reported answers in a survey. Furthermore, mobility reflects actual behavior whereas surveys are often concernd with the intent of behavior change, even though it is well known that the intent to change behavior does not fully predict actual behavior change \cite{psychology_intent_behavior}. 

Not only is mobility data widely available, but also already a tool in modern epidemiology. The structure of air transportation networks has been used to predict the spread of pathogens between countries already more than a decade ago \cite{brockmann_air_transport}. More recently, within-country mobility that is measured, e.g., using mobile phones, has become a common tool during the COVID-19 pandemic motivated by its potential to explain or predict infection dynamics \cite{lancet_mobility_infection_dynamics,ncomms_mobility_infection_dynamics}, even though the relationship between mobility and infection dynamics is non-trivial on longer time scales \cite{npj_mobility_infection_dynamics,mobility_transfer_entropy}. 

Furthermore, some studies already analyzed possible causes of mobility change \cite{brockmann_mobility_networks,npj_npi_mobility_effect,scirep_npi_mobility_effect,colizza_mobility_effects_france} and its evolution over time \cite{santana_mobility_changes_space_time}. However, these studies typically did not provide mechanistic models for mobility change but rather analyzed if there are relations at all, e.g., between \acp{GR} and mobility reduction.

On the other hand, there is also a lot of work that incorporates behavior change as a response to infection dynamics into epidemiological models \cite{funk_behavior_review}. This has been often done in a more theoretical fashion \cite{perra_characterization_of_behavior_disease_models,game_theory_behav_model,pnas_behavior_model,ieee_game_theory_vaccines,csf_model_vol_quarantine} where resulting infection dynamics are analyzed. Often, these models are not validated with real-world data, again probably due to a lack of behavior data. Thus, some authors instead validated their models with reported case numbers \cite{korea_behav_model,sci_rep_paper}. However, as argued in \cite{durham_hbm}, this approach suffers from a generally time-varying under-reporting factor and neglects the fact that infection dynamics are not solely influenced by a single \ac{PB}, but also, e.g., by other \acp{NPI}, vaccinations \cite{ncomms_vaccinations_npis_europe}, and seasonality \cite{gavenvciak2022seasonal}. 

A few exceptions exist, where detailed behavior models were proposed and also validated, e.g., in \cite{durham_hbm}, a mathematical interpretation of the \ac{HBM} \cite{hbm_definition} was developed and validated with data about mask-wearing during the 2003 SARS epidemic in Hong Kong. The effect of risk-perception on the 2009 H1N1 influenza dynamics was investigated in \cite{poletti_risk_perception_h1n1}. There, an extended \ac{SIR} model was developed and fitted not only to epidemiological data but also the behavior component compared to data about purchases of antivirals. The TELL ME-project is a cautionary tale about the necessity of validating behavioral models. Even though experts considered its results ''realistic'', a comparison of its outputs to survey data revealed stark differences to empirical results \cite{tell_me}.

In this paper, we extend the literature by introducing an \ac{ODE}-based behavioral model that is validated with aggregate mobility data from nine European countries between March 01st and December 31st, 2020. This model can reproduce the key trends of the mobility change in this period despite large inter-country differences. The key element is a time-dependent risk perception. 

We couple our behavioral model to an \ac{SIR} model and demonstrate that this can lead to complex infection dynamics, including multiple peaks of infection that resemble realistic infection patterns. Through a comprehensive analytical treatment, we identify two distinct regimes for the final number of susceptibles depending on the behavioral and epidemiological parameters. We provide a theoretical explanation for the emergence of these regimes.

The remainder of this paper is organized as follows: In Sec. \ref{sec:time_variant_behavior_model}, we explain the psychological motivations behind our model and introduce its mathematical formulation. In Sec. \ref{sec:behavior_model_validation}, we validate this behavioral model with mobility data before coupling it with an \ac{SIR} model in Sec. \ref{sec:coupled_model_introduction}. Finally, we provide an analysis of the resulting model dynamics in Sec. \ref{sec:coupled_model_analysis} and summarize our main findings and provide an outlook about potential extensions in Sec. \ref{sec:conclusion}.

\section{Behavior Model: Introduction}
\label{sec:time_variant_behavior_model}
We devise a simple, psychologically motivated model with time-variant weighing of the reported number of cases.

\subsection{Motivation}
\label{sec:time_variant_model_motivation}
Similar to Durham et al. \cite{durham_hbm}, we make use of the \ac{HBM} as foundation for our behavior model. The adoption of \ac{PB} is assumed to depend on the four factors (1) perceived likelihood of infection (\textit{susceptibility}), (2) perceived likelihood of severe disease if infected (\textit{severity}), (3) perceived benefits of, and (4) perceived barriers against adopting the corresponding behavior. 

Empirical psychological research supports this theoretical framework, i.e., that higher perceived susceptibility, higher perceived severity, and higher perceived \ac{PB} efficacy are relevant predictors of higher \ac{PB} adoption \cite{bish_michie_health_behavior_review}. 

A key insight about the \ac{HBM} is the emphasis on the \textit{perception} of these factors. So, even if the actual severity of a disease typically stays constant over time\footnote{Of course, the introduction of vaccines or increasing population immunity can reduce the actual severity. However, we focus on a time-period where these factors can be neglected.}, the perceived severity can change. The perception, and in turn adoption of \ac{PB}, can change, e.g., due to media consumption \cite{scopelliti2021tv,muniz2020media_risk_perception}, (maladaptive) coping mechanisms \cite{pilch_malcoping}, infodemics, or simply the fact that evidence about a novel disease has to be gathered over time.

Similarly, the perceived likelihood of infection might also differ from the actual likelihood of infection. One example is having a time-varying under-reporting factor. However, even the reported number of cases can be interpreted differently as time progresses, simply because humans might get used to higher infection levels and therefore reduce their reaction to them. Over time, individuals might notice that they actually have not got infected and therefore higher reported case numbers are necessary to trigger the adoption of \ac{PB}.

Furthermore, the perceived efficacy of \ac{PB} might decrease over time as shown in \cite{shiloh_behavior_adherence}. There, the authors found a high perceived efficacy early in the COVID-19 pandemic and argued that these high expectations might not have been met, causing a decrease of the perceived efficacy later on. 

Finally, the perceived barriers can reduce the adoption of \ac{PB}, e.g., the higher the social, economic, or emotional cost associated with a certain \ac{PB}, the larger the tendency to stop adopting it. 

\subsection{Model Definition}
We propose a mathematical model for the evolution of the fraction of individuals adopting \ac{PB}. Specifically, we argue that individuals either adopt \ac{PB} or not. 

In order to operationalize the insights from Section \ref{sec:time_variant_model_motivation}, we propose the \ac{ODE} 
\begin{equation}
	\deriv{p(t)}{t} = \left[ f_{0} \cdot \left(t+1\right)^a \cdot c(t) \right]\cdot \left[1-p(t)\right] + g_{0} \cdot p(t),
\end{equation}
where $f_{0},g_{0}\geq0$ and $a\leq0$ are the model parameters. $p(t) \in [0,1]$ and $c(t)$ denote the fraction of the population that adopts \ac{PB} and the reported number of cases on day $t$, respectively.

In this model, $f_{0}$ governs how ''sensitive'' the population is to the reported number of cases. The term $(t+1)^{a}$ is the time-varying factor that weighs $c(t)$. For $a<0$, the rate to adopt \ac{PB} decreases over time for a fixed value of $c(t)$. 

To interpret this expression, we first note that the uptake of \ac{PB} requires that one feels \textit{susceptible} to a \textit{severe} disease and assumes that the respective \ac{PB} is \textit{effective} in reducing their risk of infection. Thus, a decrease of any of these factors reduces the adoption rate of \ac{PB}. Hence, the time-variant factor could be seen as a decrease of a single factor or as a decrease of several factors. For example, one could argue that the perceived severity and efficacy stay constant over time, but for a given reported prevalence, the perceived susceptibility reduces. Or, the perceived susceptibility could stay constant, but both the perceived severity and \ac{PB} efficacy go down. 

Finally, $g_{0}$ governs the tendency to stop adopting \ac{PB} and is related to the (perceived) cost associated with \ac{PB}\footnote{Note that one could also argue that this cost increases over time. Then, similar model dynamics would be expected as when assuming decreasing perceived susceptibility/severity. However, a - generally unbounded - increasing function poses numerical issues and is thus not considered in this paper.}. 

\section{Behavior Model: Validation}
\label{sec:behavior_model_validation}
In this section, we show that our behavior model is able to explain key trends in real-world mobility data from various European countries, despite its very low complexity and the long time-frame under consideration. 

\subsection{Motivation}
While many epidemic models incorporating a behavioral component haven been proposed, only few of them have been validated with real-world data. As explained in the introduction, mobility data can be used as a proxy for behavior change to alleviate the issue of not being able to validate behavior models due to a lack of survey data over the relevant time-frames. 

\subsection{Data Sources and Fitting Procedure}
Google provided aggregate data about the mobility change in countries during the COVID-19 pandemic. Specifically, for certain categories of mobility, in our case \textit{retail and recreation}, the relative mobility change defined as
\begin{equation}
	\mchangeobs(t) = \frac{\mobs(t)}{\mobsbaseline(t)} - 1 \label{eq:behavior_model_validation_mobility_change_definition}.
\end{equation}
is provided on a daily level. Here, $\mobs(t)$ and $\mobsbaseline(t)$ denote the observed amount of mobility on day $t$ and the baseline amount of mobility, meaning the expected mobility in absence of any behavior change, on day $t$, respectively. Thus, a mobility reduction is indicated by negative values of $\mchangeobs(t)$.

For simplicity, we assume that an individual either is mobile or is not, i.e., we would expect no mobility at all if $p(t)=1$, i.e., if everybody adopts \ac{PB}. Therefore, we can calibrate the model parameters $f_{0}$, $a$, and $g_{0}$ by minimizing
\begin{equation}
	\mathrm{RMSE} =  \sqrt{ \frac{1}{\tobs} \sum_{t=0}^{\tobs-1} \left(p(t) - \ReLU{ -\mchangeobs(t) }  \right)^2 }, \label{eq:rmse}
\end{equation}
where $\tobs$ denotes the number of considered days and $\ReLU{\cdot}$ the rectified linear unit function.

For a detailed description of the used data and fitting procedure, we refer to \ref{sec:appendix_behav_model_validation_data_methods}. 

\subsection{Results}
\begin{figure*}
    \centering
    \includegraphics[width=0.7\textwidth]{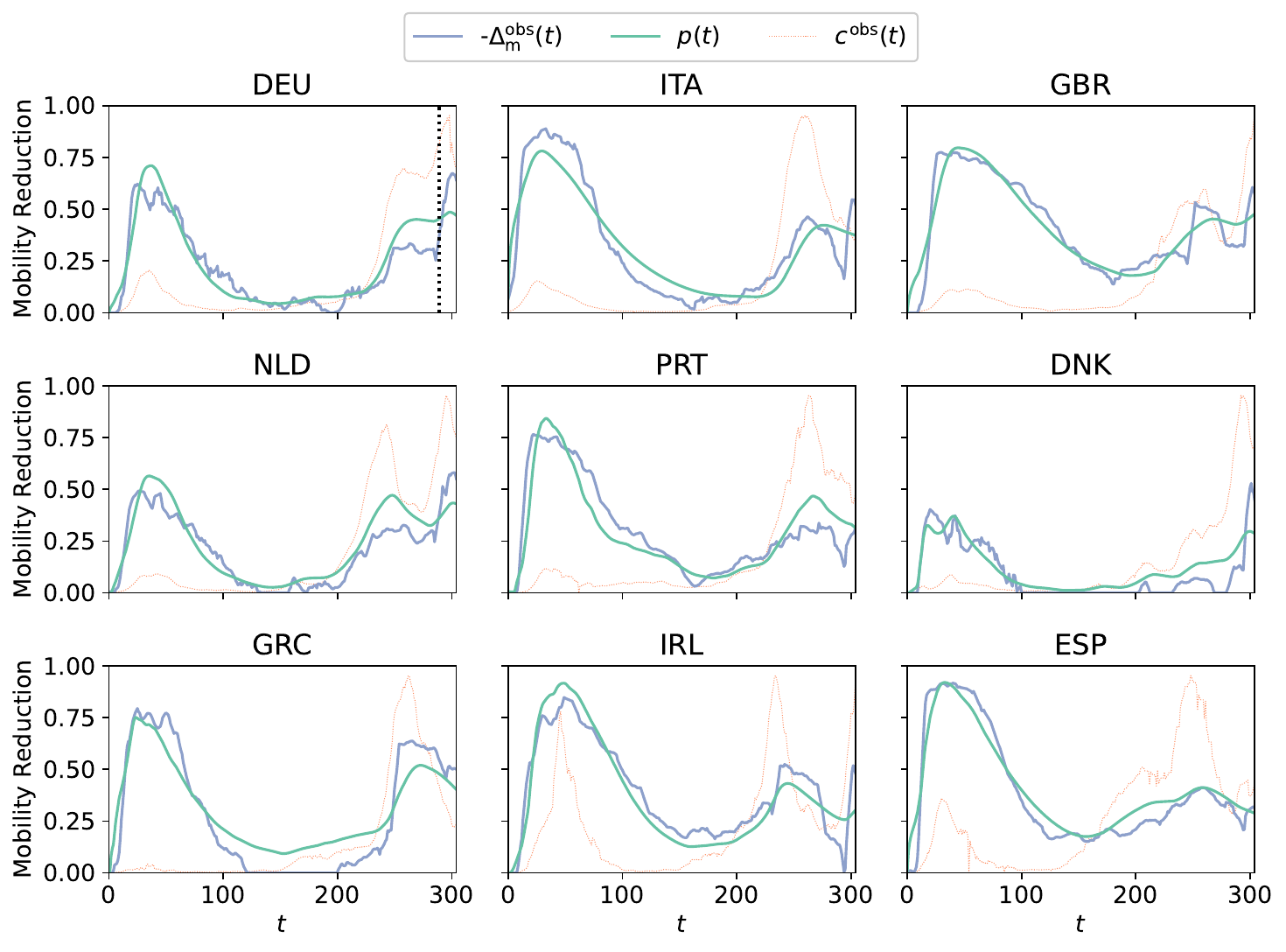}
    \caption{Observed mobility reduction (blue), corresponding model fits (green) and normalized smoothed number of reported cases (red) for nine European countries between Mar. 1st, 2020, and Dec. 31st, 2020.}
    \label{fig:behav_model_model_validation}
\end{figure*}
In Fig.~\ref{fig:behav_model_model_validation}, we show the observed mobility reduction $-\mchangeobs(t)$ in nine European countries from March 01st to Dec. 31st 2020, the fraction of individuals adopting \ac{PB} $p(t)$, and the normalized reported number of cases $c(t)$ in blue, green, and red, respectively. The inferred parameters are reported in \ref{sec:appendix_behav_model_sensitivity}, where we also show the parameters and fitting errors for the alternative end dates of Nov. 30th 2020 and Jan. 31st 2021, respectively. 

As a first observation we note the large inter-country differences of both $c(t)$ and $\mchangeobs(t)$. Yet, some general patterns emerge. In all countries, there is a large mobility reduction early on in the pandemic. Then, over time the mobility level goes back to normal. Typically, only when a pronounced second wave of infections occurs, mobility is reduced again. However, the height and duration of mobility reductions differ considerably and also whether the mobility reduction is larger during the first or second wave of infections.

Despite these large variations, the mobility reduction is well captured by our model in all countries. Especially, during the first wave of infections, the mobility patterns are almost exactly reproduced. During the second wave, the general trend of having a considerable mobility reduction is again captured. However, some specifics are not described by the model. 

For example, the mobility in Germany is additionally reduced around $t=290$ despite an earlier reduction around $t=260$. $p(t)$ remedies this by assuming a higher initial mobility reduction and only a slight subsequent increase in \ac{PB}. Indeed, it is not surprising that our model cannot reproduce these exact mobility patterns: The second mobility reduction happened due to the introduction of additional government regulations on Dec. 16, 2020 \cite{perumal2022impact}. For reference, we indicated this date by the dotted black line in Fig.~\ref{fig:behav_model_model_validation}. However, \acp{GR} are not explicitly captured by our model. While \acp{GR} are also typically in response to higher values of $c(t)$, their switch-like effect on $\mchangeobs(t)$ cannot be captured by our continuous model without signficiant extensions that would also introduce a number of new parameters.

Therefore, the observation that our model can reproduce the initial mobility change much better than at later points in time motivates the hypothesis that the role of \ac{VPB} is more relevant during the onset of the pandemic compared to when individuals are already used to it. Note that this fits well with our model assumption that the perceived severity/susceptibility reduces over time.

\section{Coupled Behavior-Disease Model: Introduction and Scenarios}
\label{sec:coupled_model_introduction}
Because our model proved useful to explain mobility change, which can be assumed as a proxy for behavior change, we go one step further and couple $p(t)$ with a model for disease spread.

Note that we do not attempt to fit the resulting infection dynamics to the reported number of cases in some country for two reasons. First, the reported number of cases are subject to a - generally time-varying - under-counting factor which would make the fitting procedure useless. Secondly, even if the true number of infections were known, or estimated, e.g., using publicly available models like \cite{ihme2021modeling}, many factors besides a single behavior influence the infection dynamics, like, seasonality \cite{gavenvciak2022seasonal}, hygiene measures \cite{restaurants_hygiene_measures}, or individuals who might adopt different \acp{PB} at different prevalences or depending on their vaccination status \cite{wambua_CoMix_study}. Also, mask-wearing, which has been shown to reduce the risk of infection considerably \cite{mask_wearing_pnas}, has not been recommended or adopted initially. Only over time, recommendations were changed and both voluntary and government-mandated usage increased.

Therefore, we do not attempt to validate the coupled model with any real-world data. Instead, we present some examples of the complex infection dynamics it can create and provide a comprehensive mathematical analysis of its long-term behavior.  

\subsection{Model Definition}
We use a modified \ac{SIR} model. Here, the population is divided into three compartments of susceptible, infectious, and recovered individuals, whose sizes are indicated by $s(t)$, $i(t)$, and $r(t)$, respectively. To facilitate theoretical analysis, we assume a normalized population, i.e., $s(t)+i(t)+r(t)=1$ $\forall t$. 

First, we note that the \ac{SIR} model can be re-written in terms of the effective reproduction number $\Rt$, i.e., 
\begin{alignat}{3}
	\deriv{s(t)}{t} &= - \frac{\Rt}{\tau} i(t) \\
	\deriv{i(t)}{t} &= + \frac{\Rt}{\tau} i(t) &- \frac{1}{\tau} i(t) \\
	\deriv{r(t)}{t} &= &+\frac{1}{\tau} i(t),
\end{alignat}
where $\tau$ denotes the average duration of infectiousness. In the standard \ac{SIR} model, $\Rt$ is given by
\begin{equation}
	\Rt = R_{0} \cdot s(t),
\end{equation}
where $R_{0}$ denotes the basic reproduction number. Now, if we have a hypothesis of how certain actions influence $\Rt$, this formulation gives us a straight-forward way to include them into an \ac{SIR} model.

In the remainder of this paper, we assume that an increase of $p(t)$ reduces the reproduction number. Specifically, if $p(t)=0$, we should obtain the same result as for the standard \ac{SIR} model. But if $p(t)=1$, we should obtain $\Rt=0$. Therefore, we use
\begin{equation}
	\Rt = R_{0} s(t) [1-p(t)]
\end{equation}
to couple our behavior model with the infection dynamics. The full model is thus given by
\begin{alignat}{3}
	\deriv{s(t)}{t} &= - \frac{R_{0} s(t) [1-p(t)]}{\tau} i(t)\label{eq:coupled_sir_s} \\
	\deriv{i(t)}{t} &= + \frac{R_{0} s(t) [1-p(t)]}{\tau} i(t) &- \frac{1}{\tau} i(t) \label{eq:coupled_sir_i} \\
	\deriv{r(t)}{t} &= &+\frac{1}{\tau} i(t) \label{eq:coupled_sir_r}\\
	\deriv{p(t)}{t} &= \left[ f_{0} \left(t+1\right)^a i(t) \right]\cdot \left[1-p(t)\right] &- g_{0} \cdot p(t) \label{eq:coupled_sir_p}
\end{alignat}
if we use $i(t)=c(t)$.

\subsection{Examples of Complex Dynamics}
\begin{figure*}
     \centering
     
     \begin{subfigure}[b]{0.45\textwidth}
         \centering
         \includegraphics[width=\textwidth]{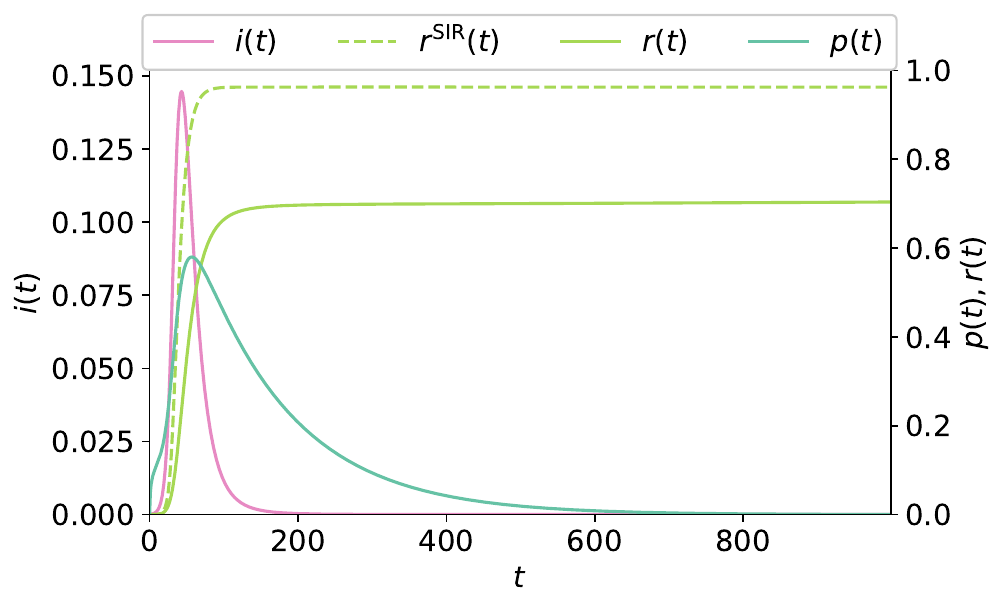}
         \vspace*{-7mm}
         \caption{Scenario 1.}
         \label{fig:coupled_model_scenario_1}
     \end{subfigure}
     \begin{subfigure}[b]{0.45\textwidth}
         \centering
         \includegraphics[width=\textwidth]{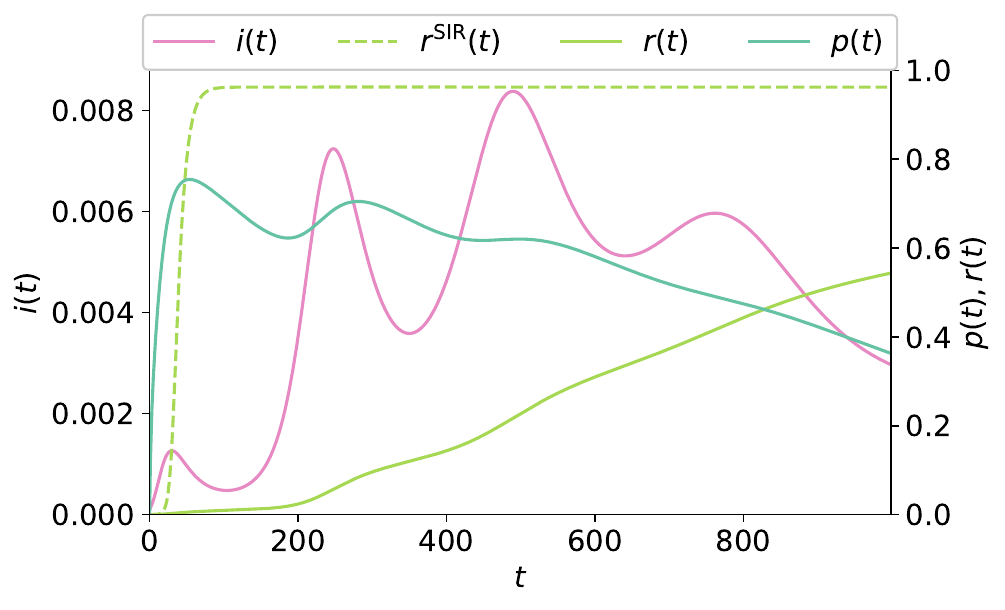}
         \vspace*{-7mm}
         \caption{Scenario 2.}
         \label{fig:coupled_model_scenario_2}
     \end{subfigure}
     \vspace*{2mm}
      
     \begin{subfigure}[b]{0.45\textwidth}
         \centering
         \includegraphics[width=\textwidth]{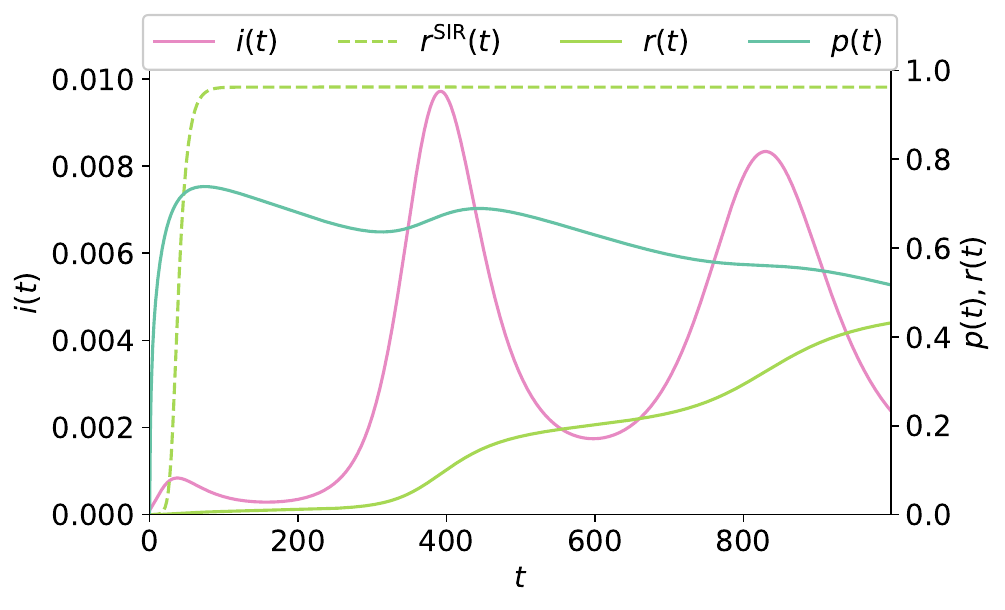}
         \vspace*{-7mm}
         \caption{Scenario 3.}
         \label{fig:coupled_model_scenario_3}
     \end{subfigure}     
     \begin{subfigure}[b]{0.45\textwidth}
         \centering
         \includegraphics[width=\textwidth]{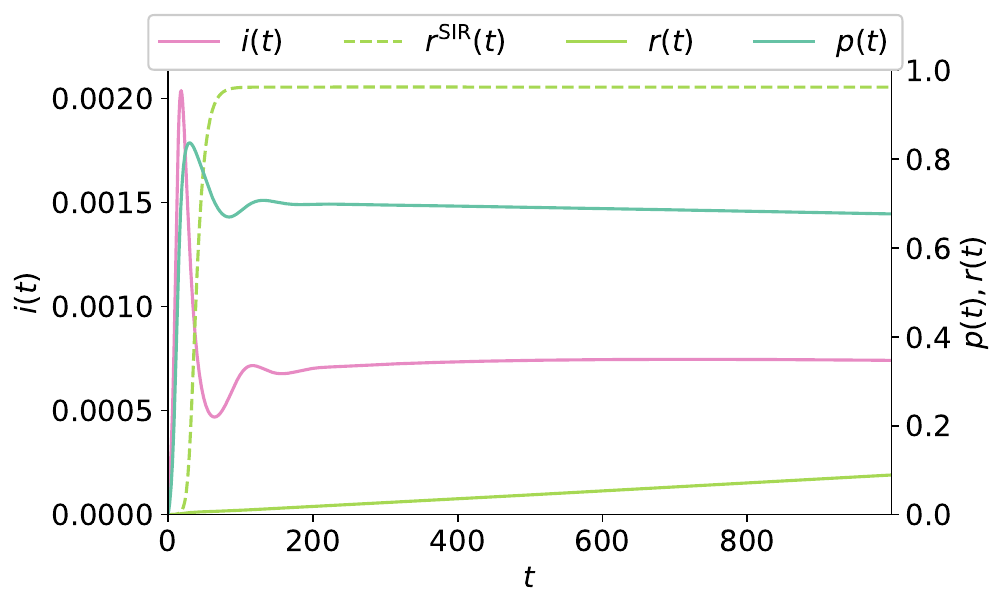}
         \vspace*{-7mm}
         \caption{Scenario 4.}
         \label{fig:coupled_model_scenario_4}
     \end{subfigure}
     \vspace*{-3mm}
        \caption{Four scenarios with the fraction of infected individuals (solid purple line), the fraction of individuals adopting \ac{PB} (solid turquoise line), and the fraction of recovered individuals (solid green line) for different behavior parameters. For reference, the fraction of recovered individuals over time in absence of any \ac{PB} (dashed green line).}
        \label{fig:coupeld_model_scenarios}
\end{figure*}
As a first step, we present some examples of the complex dynamics that can be created by our model. Four scenarios are shown in Fig.~\ref{fig:coupeld_model_scenarios}, with the corresponding parameters reported in \ref{sec:appendix_coupled_model_parameters_scenarios}. Here, $i(t)$, $r(t)$, and $p(t)$ are shown by the solid pink, green, and turquoise lines, respectively. For reference, we also show the fraction of recovered individuals that would be obtained in the standard \ac{SIR} model, $\rsir(t)$, by the dashed green line. 

The first scenario is rather similar to the standard \ac{SIR} model with a single big wave of infections. Here, $p(t)$ increases too slowly to stop the infections. Yet, through comparison of $r(t)$ and $\rsir(t)$ we can see that the adoption of \ac{PB} still reduces the prevalence, s.t., the total number of infected individuals is much lower compared to the case without any \ac{PB}.

In the second scenario, multiple waves of infections occur. There is a very fast initial increase of $p(t)$ which keeps the first wave extremely low. But over time, $p(t)$ goes down again which enables the emergence of a second wave that is stopped by an increase of $p(t)$ in response to the higher infection levels. Later, $p(t)$ goes down again and causes the emergence of a third wave of infections. After that, $p(t)$ reduces further. However, now, a considerable amount of population immunity has built up, s.t., the peak heights of the waves reduce again. 
 
The third scenario is another one with multiple waves of infection. Similar to the previous scenario, there is a very steep increase of $p(t)$ in the beginning. Again, the peak height of infection waves is determined by the interplay between \ac{PB} and population immunity. 

The fourth scenario is significantly different from the other three as an initial peak value is followed - after dampened oscillations - by approximately constant values of $p(t)$ and $i(t)$.

\section{Coupled Behavior-Disease Model: Analysis}
\label{sec:coupled_model_analysis}
Now that we have seen the wide variety of infection dynamics resulting from time-variant risk perception, we provide some analytical insights into the model dynamics. 

First, we look at \eqref{eq:coupled_sir_p} and note that for $a<0$ we have $f_{0} \left(t+1\right)^a i(t)=0$ for $t \rightarrow \infty$ because $i(t)$ can be bounded from above by some constant $\imax$, e.g., $\imax=1$. Therefore, for any $g_{0}>0$, we have $p(t)=0$ for $t \rightarrow \infty$. By looking at \eqref{eq:coupled_sir_s}-\eqref{eq:coupled_sir_r}, we see that $i(t)=0$ has to hold, s.t., $s(t)$, $i(t)$, and $r(t)$ do not change anymore.

From this, we know the fate of our model for $t \rightarrow \infty$: Eventually nobody will adopt \ac{PB} anymore and there won't be any infections\footnote{However, even in models where re-infections are possible, like in the SIRS model, nobody would adopt \ac{PB} for $t \rightarrow \infty$ if $a<0$.}. However, how long it takes to reach this state and the final values of $s(t)$ and $r(t)$ are generally unknown.

\subsection{(Approximate) Fixed Point in the Low-Infection Limit}
If $\Rt\approx1$ and $i(t)$ is very small, it can take a very long time until this final state is reached. In fact, it can happen that $r(t)$ increases slow enough such that population immunity has no relevant effect for a very long time. In this case, one can approximate $s(t)\approx1$. If additionally $a \approx 0$, i.e., the risk-awareness is almost time-invariant, the model dynamics are approximately captured by 
\begin{alignat}{3}
	\deriv{i(t)}{t} &= \frac{R_{0} [1-p(t)]}{\tau} i(t) &- \frac{1}{\tau} i(t) \label{eq:coupled_sir_low_infections_i} \\
	\deriv{p(t)}{t} &= f_{0} i(t) \cdot \left[1-p(t)\right] &- g_{0} \cdot p(t). \label{eq:coupled_sir_low_infections_p}
\end{alignat}
In this setting, $i(t)$ and $p(t)$ are, for large values of $t$, almost constant and given by
\begin{equation}
	\iinfty = \frac{g_0}{f_0} (R_0-1)
\end{equation}
and
\begin{equation}
	\pinfty = 1-\frac{1}{R_0},
\end{equation}
respectively. For the full derivation and stability analysis, we refer to \ref{sec:appendix_coupled_model_derivations_fixed_points}. If we use the same parameters as for Scenario 4, we obtain $\iinfty  \approx 4.3 \cdot 10^{-4}$ and $\pinfty \approx 0.7$. These values are indeed of the same order of magnitude as the actual values of $i(t)$ and $p(t)$ that are displayed in Fig.~\ref{fig:coupled_model_scenario_4}.

\subsection{Regimes for Long-Term Infection Dynamics}
\begin{figure}
    \centering
    \includegraphics[width=0.5\textwidth]{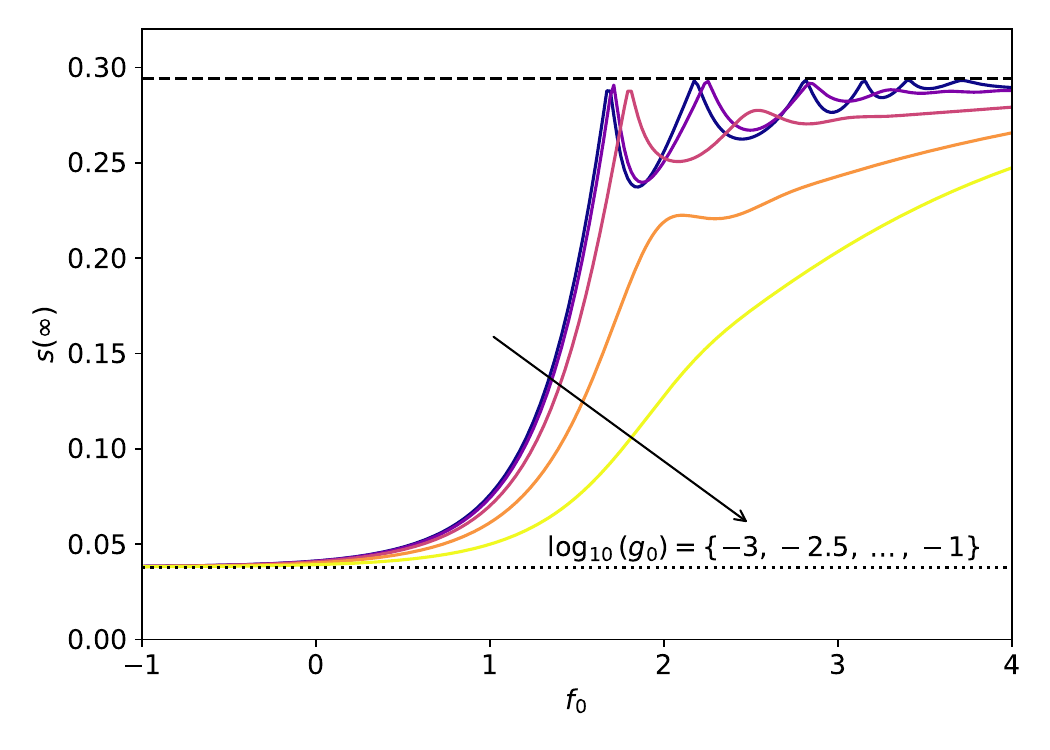}
    \vspace*{-10mm}
    \caption{The fraction of susceptible individuals after $t=5 \cdot 10^4$ days as a function of $f_0$. The results are reported for different values of $\logten{g_0}$ which are indicated by increasingly bright colors as $\logten{g_0}$ increases.}
    \label{fig:coupled_model_analysis_sfinal_ensemble}
\end{figure}
Generally $a\neq0$ and $i(t)$ can become large enough such that the previous analysis of the low-infection limit is not valid. Also, as we have shown in Fig.~\ref{fig:coupeld_model_scenarios}, the full behavior-disease model can produce a wide variety of infection dynamics that are hard to predict and understand analytically. 

However, the final value of suscepetibles $s(\infty)$ might be more predictable. Motivated by this, we plot $s(\infty)$ over $\logten{f_0}$ for different values of $g_0$ in Fig.~\ref{fig:coupled_model_analysis_sfinal_ensemble}. The corresponding simulation parameters are reported in \ref{sec:appendix_coupled_model_f0_ensemble}. 

One can distinguish two regimes for the final number of susceptibles: For smaller values of $f_0$, $s(\infty) \approx \ssir{\infty} \approx 0.04$ and for large values of $f_0$, $s(\infty) \approx \frac{1}{3.4} \approx 0.3$. In-between there is a transition between these regimes. First, we explain these values and show how the regime can be predicted based on the initial model dynamics. Finally, we explain the non-smooth dynamics that can be observed for smaller values of $g_0$.

\subsubsection{Explaining the Regime Values}
As described earlier, for any $a < 0$, $p(t)$ will eventually be zero. Therefore, only population immunity can bring $\Rt$ below one in order to exhaust the epidemic. The necessary condition for $\Rt<1$, if $p(t)=0$, is 
\begin{equation}
	s(t) < \frac{1}{R_0}.
\end{equation}
This upper bound for $s(\infty)$ can be only achieved if $i(t)$ stays small. To illustrate this, consider a wave of infections where $i(t)$ has become large, e.g., $i(t) \approx \frac{1}{5}$. Now, even if $\Rt<1$ due to population immunity, the wave has still \textit{momentum}, i.e., many more individuals are still getting infected even though the reproduction number is below one. 

This is exactly what happens in the standard \ac{SIR} model without \ac{PB}. Indeed, for smaller values of $f_0$, $s(\infty) \approx \ssir(\infty)$, where $\ssir(\infty)$ denotes the final number of susceptibles in the standard \ac{SIR} model. For reference, we plotted this as the dotted black line in Fig.~\ref{fig:coupled_model_analysis_sfinal_ensemble}. Indeed, this is the value of the regime with fewer final susceptibles.

On the other hand, if $i(t) \approx 0$, $s(\infty)$ can actually reach the upper bound of $\frac{1}{R_0}$. For reference, we plotted this as the dashed black line in Fig.~\ref{fig:coupled_model_analysis_sfinal_ensemble}. Indeed, this is the value of the regime with more final susceptibles. 

\subsubsection{Predicting the Regime for Long-Term Infection Dynamics}
Let us check if we can predict in which regime one ends up for given behavior and disease parameters without running the full simulations. 

To this end, we ask whether $p(t)$ is large enough, s.t., \ac{PB} can significantly decrease $\Rt$ before  too many individuals got infected in the first wave. First, we observe that $i(t)$ can be approximated for $t \rightarrow 0$, i.e., when $p(t) \approx 0$ and $s(t)\approx$1 by
\begin{equation}
	i(t) \approx i_0 \exp\left( \frac{R_0-1}{\tau}t \right) = i_0 \exp\left(\lambda_0 t\right)\label{eq:coupled_model_regime_prediction_i_exp},
\end{equation}
where $i_0=i(0)$. Similarly, we simplify \eqref{eq:coupled_sir_p} by assuming $g_0 \approx 0$ and $p(t) \approx 0$ for $t \rightarrow 0$. In this case, $\deriv{p(t)}{t}$ depends only on $t$ and $i(t)$. Thus, if we have a closed-form expression for $i(t)$ like our approximation in \eqref{eq:coupled_model_regime_prediction_i_exp}, we can obtain $p(t)$ by solving the integral
\begin{equation}
	p(t) = \int f_0 i(t) (t+1)^{a} \dt \label{eq:coupled_model_regime_prediction_p_approx}
\end{equation}
with the initial condition $p(0) = p_0$.

Then, we can ask whether $i(t)$ would reach a value $\istar>0$ before or after \ac{PB} has a relevant effect, i.e., when $p(t)$ reaches some value $\pstar = \beta \cdot \istar$ with $\beta \in (0,1]$. 

\begin{figure}
    \centering
    \includegraphics[width=0.5\textwidth]{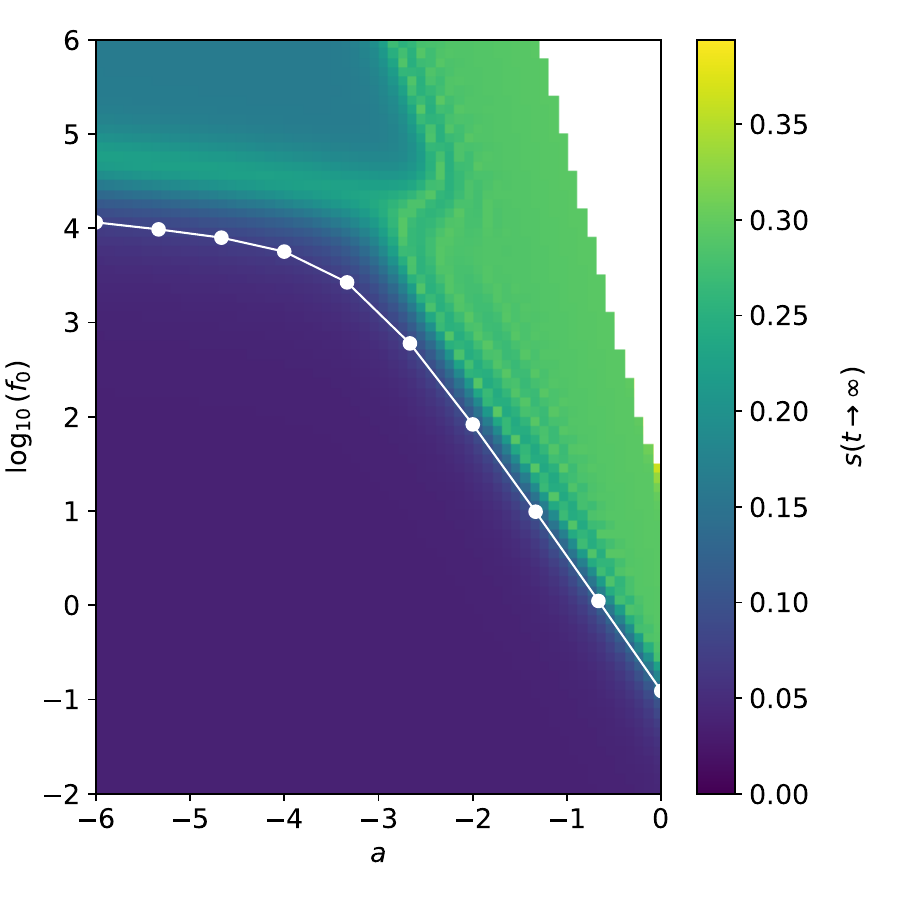}
    \vspace*{-7mm}
    \caption{The fraction of susceptible individuals as a function of $f_0$ and $a$ for exemplary behavior and disease parameters after $t=5 \cdot 10^4$ days. We indicated the approximated transition between both regime values by the solid white line. The white area indicates parameter combinations where the simulations have not yet converged after $5 \cdot 10^4$ days.}
    \label{fig:coupled_model_analysis_f0-a-plane_sfinal}
\end{figure}

To validate this idea, we run simulations for different values of $\logten{f_0}$ and $a$ with parameters as described in \ref{sec:appendix_coupled_model_parameters_final_susceptibles}. We then compute the curve in the $a$-$f_0$-plane where $p(t)$ reaches $\pstar = \frac{1}{2} \istar =
\frac{1}{4}$ at the same time as $i(t)$ reaches $\istar=\frac{1}{2}$. The procedure to obtain this curve and additional sensitivity analyses are reported in \ref{sec:appendix_coupled_model_derivations_regime_bounds}. In Fig.~\ref{fig:coupled_model_analysis_f0-a-plane_sfinal}, we indicate $s(\infty)$ by the color and draw the obtained critical curve in white. 

For large values of $\logten{f_0}$ and small values of $a$, the simulations might take an extremely long time to converge. For example, for $a=0$ and $\logten{f_0}=5$, the low-infection limit holds and we would obtain $\iinfty=\frac{10^{-2.5}}{10^5} (3.4-1) \approx 10^{-7}$, i.e., approximately $10^7$ time steps would be necessary until all individuals got infected. In order to reduce the computational load, we use $\tsim = 5 \cdot 10^4$ time steps and do not report $s(\tsim) \approx s(\infty)$ if $i(\tsim)>\frac{1}{10} \cdot \iinfty$.

One can clearly distinguish the two regimes in the parameter space. If $\logten{f_0}$ is large and $|a|$ is small, $s(\infty) \approx \frac{1}{R_0}$, while $s(\infty) \approx \ssir(\infty)$ if $\logten{f_0}$ is small and $|a|$ is large. The border between both regimes is predicted remarkably well by our approach. Analyzing whether $p(t)$ has a significant effect in the beginning enables interestingly enough insights about the long-term behavior of $s(t)$. 

\subsubsection{Explaining the Non-Smooth Dependency on $f_0$}
\label{sec:coupled_model_final_regimes_non_smoothness_explaination}

\begin{figure}
    \centering
    \includegraphics[width=0.5\textwidth]{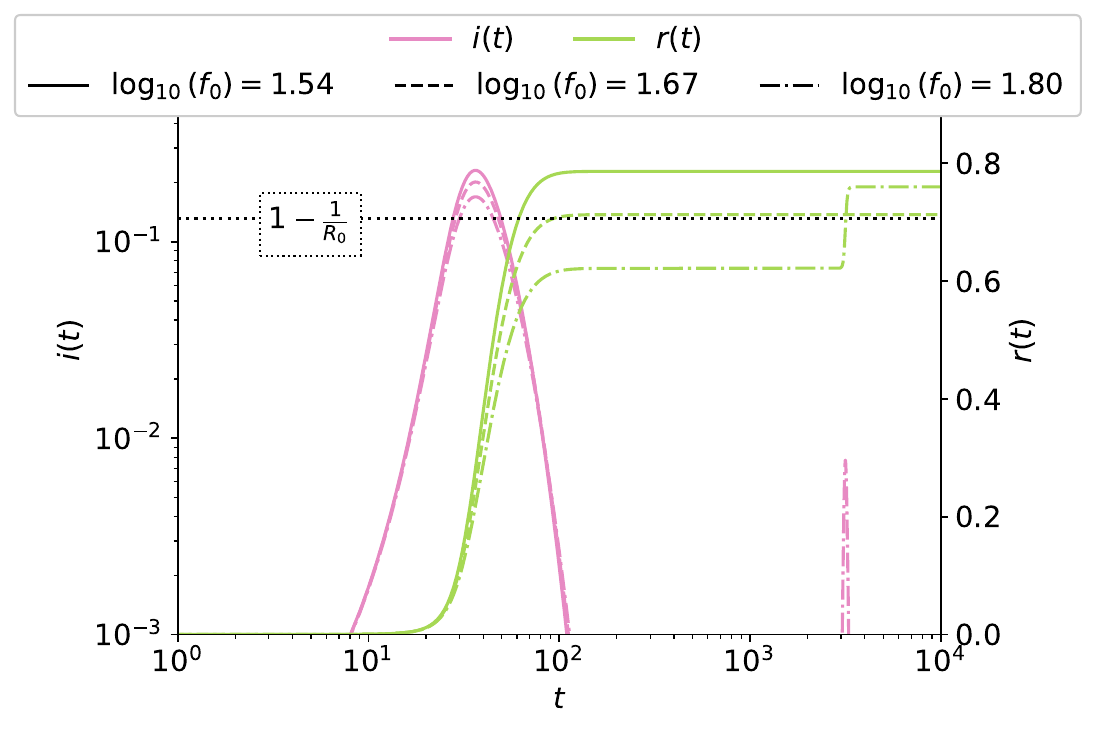}
    \vspace{-7mm}
    \caption{The fraction of infected (purple) and recovered (green) individuals for scenarios with different $f_0$. The non-monotonous behavior in Fig.~\ref{fig:coupled_model_analysis_f0-a-plane_sfinal} can be explained by the emergence of a second wave.}
    \label{fig:coupled_model_analysis_non_smoothness_explanation}
\end{figure}

As a final contribution of this paper, we look at the non-smooth dependency of $s(\infty)$ on $f_0$ that can be observed in Fig.~\ref{fig:coupled_model_analysis_sfinal_ensemble}. This is a surprising behavior because we would expect that a higher risk perception enables more individuals to escape infection. While this would be true for time-invariant perception, i.e., $a=0$, the time-dependency causes more complex dynamics.

Indeed, as $f_0$ increases, the (initial) peak height of infections decreases and fewer people get infected in the first wave. However, if $f_0$ increases further, a small second wave can emerge, if fewer than $1-\frac{1}{R_0}$ individuals got infected in the first wave, causing $s(\infty)\ll\frac{1}{R_0}$. 

We illustrate this by plotting the time-evolution of $i(t)$ and $r(t)$ for $\logten{f_0} \in \{1.54, 1.67, 1.80\}$ in Fig.~\ref{fig:coupled_model_analysis_non_smoothness_explanation} with the same parameters as used for Fig.~\ref{fig:coupled_model_analysis_sfinal_ensemble} with $\logten{g_0}=-3.0$. Indeed, the peak height of the first wave reduces with increasing values of $\logten{f_0}$ but for $\logten{f_0}=1.80$, a second wave can emerge causing more overall infections compared to the scenario with $\logten{f_0}=1.67$. We also plotted $1-\frac{1}{R_0}$ into Fig.~\ref{fig:coupled_model_analysis_non_smoothness_explanation}, s.t., we can compare $r(t)$ to the critical value for which population immunity prevents $R(t)>1$. Indeed, only in the scenario with $\logten{f_0}=1.80$, enough susceptibles are left after the first wave to cause a second wave.

\section{Conclusion}
\label{sec:conclusion}
In this paper, we proposed a simple model to describe the behavioral response to the outbreak of an emerging disease. Our model is motivated by the \acl{HBM}, a commonly used concept from health psychology that assumes that the adoption of \acl{PB} depends on the perceived severity of a disease and the perceived likelihood of infection. Therefore, individuals adopt \acl{PB} in response to the reported disease prevalence in our model. However, over time the perceived risk associated with the reported prevalence goes down. 

We validated the proposed behavior model with real-world mobility data during the COVID-19 pandemic in 2020. Despite its simplicity and its lack of explicit consideration of \aclp{GR}, the major trends of the observed mobility reduction are reproduced for different European countries. This indicates that time-varying risk perception facilitates understanding the behavioral response to the outbreak of a novel disease over long periods of time.

Additionally, we analyzed what happens if the behavior model is coupled with an \acl{SIR} model for infection dynamics. We demonstrate the emergence of complex dynamics, including multiple waves of infections. We identified two regimes for the final number of susceptibles: one regime with similar values to the default \acl{SIR} model and the other with similar values close to $\frac{1}{R_0}$, where $R_0$ is the basic reproduction number. Through detailed analysis, we derived a bound to predict the transition between both regimes depending on whether the adoption of \acl{PB} is fast enough. 

Finally, we observed that the model exhibits a non-monotonous relationship of increasing risk-perception and the final number of susceptibles. This can be explained through the emergence of subsequent infection waves as a consequence of the complex interplay of \acl{PB}, population immunity, and changes in risk-perception over time.

In future work, one could extend the behavior model to incorporate \aclp{GR} and other factors like the availability of vaccines or the emergence of \acl{VoC}. This could enable an improved model fit over longer time-frames. Furthermore, one could incorporate the behavior model in a more realistic model for disease spread that captures, e.g., the possibility of re-infections or non-homogenous mixing of different population groups. 

\section*{Acknowledgement}
Bastian Heinlein would like to thank Reinhard German and Anatoli Djanatliev for their support to enable his research stay with Manlio De Domenico. Furthermore, he would like to thank Timo Jakumeit and Hannah Stephan for their valuable comments on the manuscript.

\bibliographystyle{elsarticle-num} 
\bibliography{literature.bib}

\appendix
\section{Behavior Model Validation: Data Sources and Methods}
\label{sec:appendix_behav_model_validation_data_methods}
\subsection{Data Sources}
To validate our behavior model, we use the smoothed number of reported cases per 100,000 inhabitants provided by \ac{OWID} \cite{owidcoronavirus} as $c(t)$. As additional pre-processing we normalize these values such that $c(t)$ has unit variance between Mar. 1st, 2020 and Feb. 01st, 2021.

We use aggregate mobility data on a country level for the category \textit{retail and recreation} as reported by Google \cite{google}. In these data, daily mobility change relative to a pre-pandemic baseline is given. However, because the raw data is subject to strong periodic weekly fluctuations, we convolve the data with a \ac{FIR} filter with impulse response $h(t) = \frac{1}{7} \sum_{\tau=0}^6 \delta(t-\tau)$, where $\delta(\cdot)$ is the Delta-Dirac function.

\subsection{ODE Simulations}
Note that we have only discrete-time data $c^{\mathrm{discrete}}[ k ],\; k \in \mathbb{Z}$ with a resolution on a daily level, but \acp{ODE}, i.e., a continuous simulation method. Therefore, we use
\begin{equation}
	c(t) = c^{\mathrm{discrete}}[ \lfloor t \rfloor ],
\end{equation}
where $\lfloor \cdot \rfloor$ denotes the floor function. Comparing $p(t)$ to mobility data is straight-forward because we can sample $p(t)$ at non-negative integer values of $t$.

The \acp{ODE} are simulated using the \verb|RK45| method of \textit{scipy} \cite{scipy} with \verb|atol| $=10^{-6}$ and \verb|rtol| $=10^{-3}$. 

\subsection{Parameter Fitting}
We use $p(0) = \max\{0,-\mchangeobs(0)\}$ as initial condition. For the fitting procedure, we use only negative values for the mobility change, i.e., we compare $p(t)$ to $\ReLU{-\mchangeobs(t)}$ because our mobility model assumes that $\mobs(t) = \mobsbaseline \cdot [1-p(t)]$ and thus can only model mobility decrease. 

We use the \textit{lmfit} wrapper \cite{newville2016lmfit} around the \textit{scipy} \cite{scipy} implementation of differential evolution \cite{storn1997differential} to solve the optimization problem of finding $\logten{f_0} \in [-1,3]$, $\logten{g_0} \in [-3,0]$, $a \in [-3,0]$ that minimize \eqref{eq:rmse}.

\section{Behavior Model Validation: Sensitivity Analysis}
\label{sec:appendix_behav_model_sensitivity}
To verify that our behavior model is a robust explanation for mobility change, we performed additional parameter inferences as sensitivity analysis. Specifically, we varied the end date of the considered time period. Otherwise, we use the same simulation and inference procedure as for the default case. 

\begin{figure*}
    \centering
    \includegraphics[width=\textwidth]{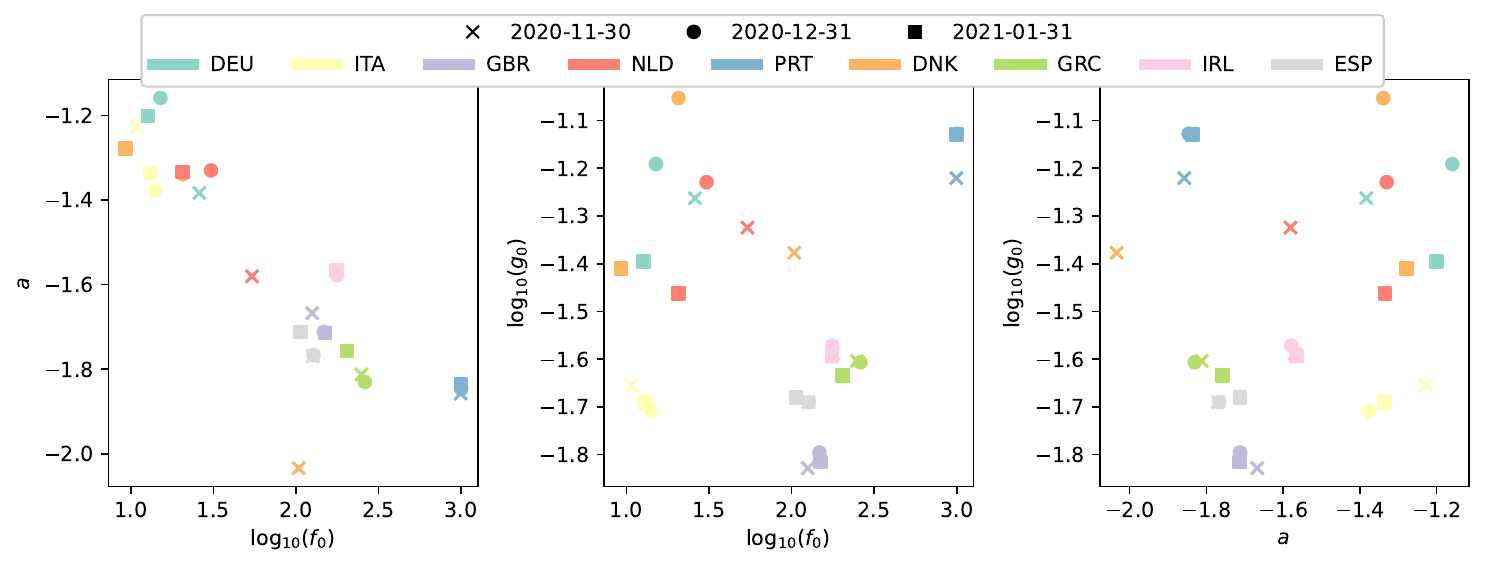}
    \caption{Best-fitting model parameters for the considered European countries (indicated by color) and varying end dates (indicated by the marker shape).}
    \label{fig:appendix_behav_model_validation_inferred_params}
\end{figure*}

In Fig.~\ref{fig:appendix_behav_model_validation_inferred_params}, we show the inferred parameters for $\logten{f_0}$, $\logten{g_0}$, and $a$ depending on the chosen time frame. The different countries are indicated by the colors and the different end dates by the marker. For most countries, the inferred parameter values do not change much depending on the end date. Denmark has the largest variations which can be explained by looking at Fig.~\ref{fig:behav_model_model_validation}. The second larger mobility reduction happens at the very end of 2020 and thus is not captured by the dataset with Dec. 31st, 2020 as end date. Additionally, the increase of mobility reduction is extremely steep despite the fact that case numbers already increased considerably before. This indicates that the introduction of \acp{GR} is an important factor here, as well. 

\begin{figure*}
    \centering
    \includegraphics[width=\textwidth]{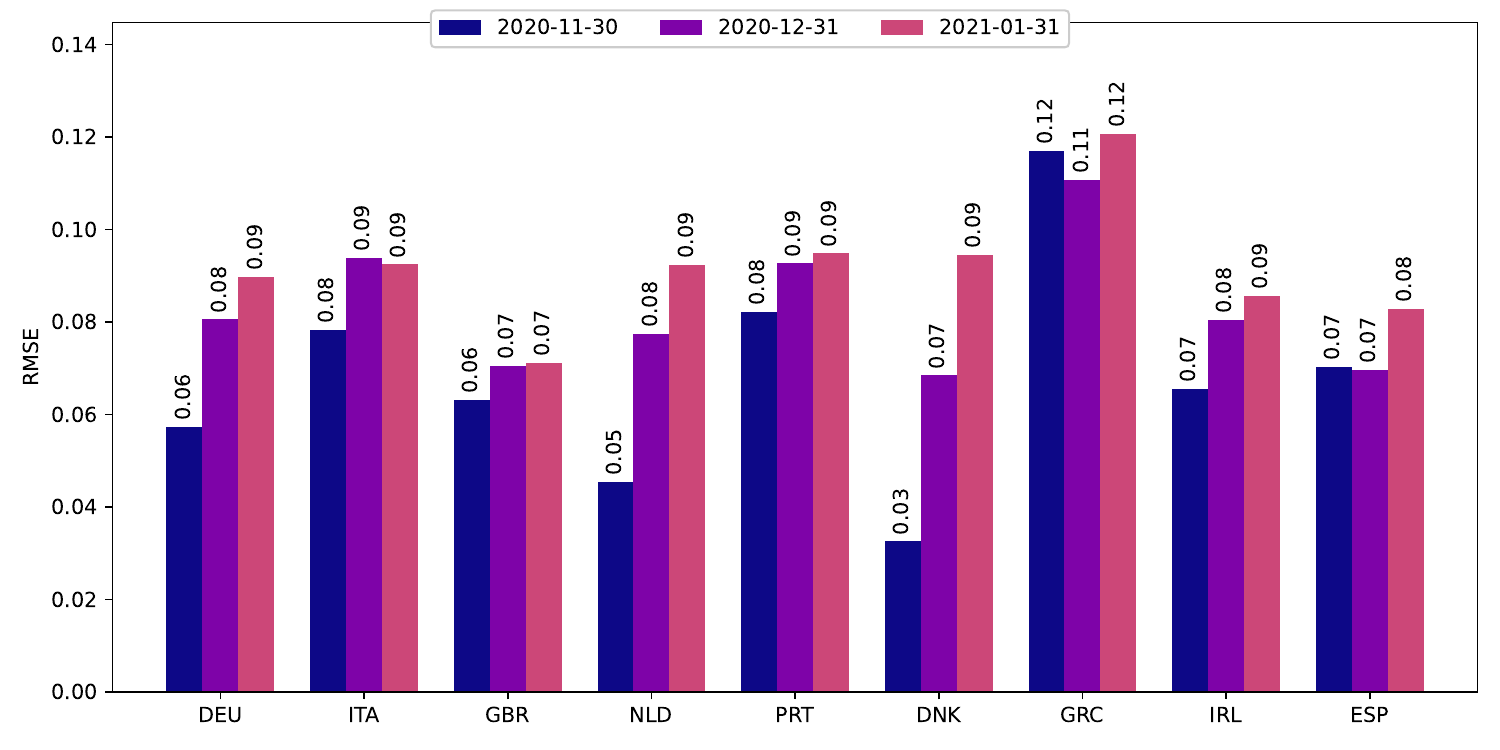}
    \caption{Errors between the estimated best-fitting model and observed mobility reduction in European countries for different end dates.}
    \label{fig:appendix_behav_model_validation_errors}
\end{figure*}

In Fig.~\ref{fig:appendix_behav_model_validation_errors}, we show the RMSE for each end date and each country. In most countries, the error increases as time increases, probably because the longer the time-frame the more ''opportunity'' for other factors besides the reported number of cases to influence the population behavior. Overall, we observe that our model inference behaves robustly.

\section{Coupled Behavior-Disease Model: Parameter Values}
\label{sec:appendix_coupled_model_parameters}

\subsection{Scenarios}
\label{sec:appendix_coupled_model_parameters_scenarios}
\begin{table}
\centering
\begin{tabular}{c || c | c | c | c | c} 
	Parameter & $\tsim$ & $R_0$ & $\tau$ & $i_0$ & $p_0$ \\ 
  	\hline\hline
 	Value & 1000 & 3.4 & 8.3  & $10^{-4}$ & $10^{-4}$\\ 
\end{tabular}
\caption{Shared parameters used in all scenarios for Fig.~\ref{fig:coupeld_model_scenarios}.}
\label{table:appendix_coupled_model_parameters_shared_among_scenarios}
\end{table}
\begin{table}
\centering
\begin{tabular}{c || c | c | c  } 
	 & $\logten{f_0}$ & $a$ & $\logten{g_0}$ \\ 
  	\hline\hline
 	Scenario 1 & 3.0 & -2.25 & -2.10  \\ 
	Scenario 2 & 3.25 &-1.30 & -2.55 \\
	Scenario 3 & 3.50 & -1.50 & -3.00 \\
	Scenario 4 & 2.00 & -0.10 & -1.75 
\end{tabular}
\caption{Scenario-specific parameters used for Fig.~\ref{fig:coupeld_model_scenarios}.}
\label{table:appendix_coupled_model_parameters_different_among_scenarios}
\end{table}
The parameters shared between the four exemplary scenarios are reported in Table \ref{table:appendix_coupled_model_parameters_shared_among_scenarios}. The values for $R_0$ and $\tau$ are taken from \cite{priesemann_science} in order to resemble values similar to the ones observed in the COVID-19 pandemic before the emergence of \ac{VoC}. In Table \ref{table:appendix_coupled_model_parameters_different_among_scenarios}, we report the behavior parameters used for the respective scenarios.

\subsection{Non-Smooth Dependency of $s(\infty)$ on $f_0$}
\label{sec:appendix_coupled_model_f0_ensemble}
\begin{table}
\centering
\begin{tabular}{c || c | c} 
	Parameter & Meaning & Value \\ 
  	\hline\hline
	$\tsim$ & Simulation Time & $5 \cdot 10^4$ \\
	\verb|atol| & Absolute tolerance for \verb|RK45| & $10^{-20}$ \\
	\verb|rtol| & Relative tolerance for \verb|RK45| & $10^{-10}$ \\
	
	\hline
	$R_0$ & Basic Reproduction Number & 3.4 \\
	$\tau$ & Mean Duration of Infectiousness & 8.3 \\
	$i_0$ & Fraction of initially infected & $10^{-4}$ \\
	$p_0$ & Fraction initially adopting \ac{PB} & $10^{-4}$ \\
	
	\hline
	
	$a$ & Forgetting Factor & -1.5 
\end{tabular}
\caption{Fixed parameter values used to create Fig.~\ref{fig:coupled_model_analysis_sfinal_ensemble}.}
\label{table:appendix_coupled_model_f0_ensemble}
\end{table}
We show the parameters used to create Fig.~\ref{fig:coupled_model_analysis_sfinal_ensemble} in Table \ref{table:appendix_coupled_model_f0_ensemble}. For each curve, we vary $\logten{f_0}$ between -1 and 4 over 250 steps. Each curve corresponds to a value of $\logten{g_0} \in \{-3.0,-2.5,-2.0,-1.5,-1.0\}$.

\subsection{$a$-$f_0$-Plane}
\label{sec:appendix_coupled_model_parameters_final_susceptibles}
\begin{table}
\centering
\begin{tabular}{c || c | c} 
	Parameter & Meaning & Value \\ 
  	\hline\hline
	$\tsim$ & Simulation Time & $5 \cdot 10^4$ \\
	\verb|atol| & Absolute tolerance for \verb|RK45| & $10^{-12}$ \\
	\verb|rtol| & Relative tolerance for \verb|RK45| & $10^{-6}$ \\
	
	\hline
	$R_0$ & Basic Reproduction Number & 3.4 \\
	$\tau$ & Mean Duration of Infectiousness & 8.3 \\
	$i_0$ & Fraction of initially infected & $10^{-4}$ \\
	$p_0$ & Fraction initially adopting \ac{PB} & $10^{-4}$ \\
	
	\hline
	
	$\logten{g_0}$ & Tendency to stop adopting \ac{PB} & -2.5 \\
	
	\hline
	
	$\istar$ & - & $\frac{1}{2}$ \\
	$\beta$ & - & $\frac{1}{2}$
\end{tabular}
\caption{Fixed parameter values used to create Fig.~\ref{fig:coupled_model_analysis_f0-a-plane_sfinal}.}
\label{table:appendix_coupled_model_f0_a_map_main_parameters}
\end{table}
We report the fixed simulation parameters for the $a$-$f_0$-plane in Table \ref{table:appendix_coupled_model_f0_a_map_main_parameters}. For the simulations we vary $\logten{f_0}$ between $-2$ and $6$ over 80 uniformly spaced steps. In the same manner, we vary $a$ between $-6$ and $0$ over 60 uniformly spaced steps.

\section{Coupled Behavior-Disease Model: Full Derivations}
\subsection{Regime Bounds}
\label{sec:appendix_coupled_model_derivations_regime_bounds}
Here, we explain how the critical curve in the $a$-$f_0$-plane can be obtained. First, we observe that the time $\tistar$ at which $i(t)=\istar$ can obtained as 
\begin{equation}
	\tistar = \frac{1}{\lambda_0} \log\left( \frac{\istar}{i_0} \right)
\end{equation}
by solving \eqref{eq:coupled_model_regime_prediction_i_exp} for $t$.

On the other hand, obtaining the time when $p(t) = \pstar$ is more difficult. First, we need an analytical expression for $p(t)$. If we insert \eqref{eq:coupled_model_regime_prediction_i_exp} into \eqref{eq:coupled_model_regime_prediction_p_approx}, we get
\begin{equation}
	p(t) = \int f_0 i_0 e^{\lambda_0 t} (t+1)^a \dt,
\end{equation}
for which no closed form integral solution exists to the best of the authors' knowledge. However, we can approximate $p(t)$ by writing $\exp(\lambda_0 t)$ as a Taylor series of $n$-th order around $t=-1$, i.e.,
\begin{equation}
	e^{\lambda_0 t} \approx \sum_{k=0}^{n} \frac{\lambda_0^k}{k!} e^{-\lambda_0} (t+1)^k.
\end{equation}
This yields
\begin{alignat}{2}
	p(t) 	&\approx \int f_0 i_0 (t+1)^a \sum_{k=0}^{n} \frac{\lambda_0^k}{k!} e^{-\lambda_0} (t+1)^k \dt + c\\
		&= f_0 i_0 e^{-\lambda_0} \sum_{k=0}^n \frac{\lambda_0^k}{k!} \int (t+1)^{a+k} \dt + c\\
		&= f_0 i_0 e^{-\lambda_0} \sum_{k=0}^n \frac{\lambda_0^k}{k!} \begin{cases} \frac{1}{a+k+1} (t+1)^{a+k+1} & \text{if }a+k \neq -1 \\ \log(1+t) & \text{if } a+k=-1\end{cases} + c.
\end{alignat}
Here, $c$ is a constant ensuring that $p(0)=p_0$.

\begin{algorithm}
    \caption{Equation Solver Algorithm}
    \begin{algorithmic}
        \Require $x_0$, $f(x)$, $\ystar$, $\niterations$
        \For{$i=1,\dots,\niterations$}
            \State $y_i = f(x_{i-1})$

            \If{$y_i < \ystar$}
            	\State $x_i \gets x_{i-1} + \stepsize$
	    \Else
	    	\State $\stepsize \gets \frac{1}{2}\stepsize$
		\State $x_i \gets x_{i-1} - \stepsize$
            \EndIf
        \EndFor
        \State\Return $x_i$
    \end{algorithmic}
    \label{alg:equation_solver}
\end{algorithm}

We cannot expect to find an analytical expression for $\tpstar$, i.e, when $p(t) = \pstar$. However, we know that $p(t)$ is a monotonously increasing function for $t \geq 0$. Therefore, we can apply Algorithm \ref{alg:equation_solver} to find $\tpstar$ with $f(t)=p(t)$, $\stepsize=10.0$, $x_0=t_0=0.0$, $\niterations=20$, and $\ystar=\pstar$.

To obtain a point on the critical curve in the $f_0-a$-plane for a given $a$, we also use Algorithm \ref{alg:equation_solver} but with $\geq$ instead of $<$ in the \verb|if|-condition. However, here $f(x)$ is $\tpstar(f_0)$ obtained as described by the previous paragraph. We thus use $x_0=-2$, $\stepsize=1.0$, $\niterations=20$, and $\ystar=\tistar$.

\subsection{Fixed Point Analysis}
\label{sec:appendix_coupled_model_derivations_fixed_points}
Here, we provide more detailed derivations for the fixed point analysis in the low infection regime.

\subsubsection{Determination of the Fixed Point}
To compute the fixed point in the low-infection limit, we insert $i(t)=\iinfty$ and $p(t)=\pinfty$  into \eqref{eq:coupled_sir_low_infections_i} and  \eqref{eq:coupled_sir_low_infections_p} and set the resulting equations to zero. First,  \eqref{eq:coupled_sir_low_infections_i} is solved straight-forwardly for $\pinfty$. Then,  \eqref{eq:coupled_sir_low_infections_p} is solved for $\iinfty$ and the previously obtained definition for $\pinfty$ is inserted.

\subsubsection{Fixed Point Stability}
To determine whether the obtained fixed point is stable, we first write the \ac{ODE} system in vector form, i.e., 
\begin{equation}
	\x = \begin{bmatrix}
		\deriv{i(t)}{t} \\
		\deriv{p(t)}{t}
	\end{bmatrix} = \begin{bmatrix}
		h_1\left( p(t), i(t) \right) \\
		h_2 \left( p(t), i(t) \right)
	\end{bmatrix}.
\end{equation}
Then, $\x_*=\begin{bmatrix}\iinfty & \pinfty \end{bmatrix}^\transpose$ is a stable fixed point if both eigenvalues of the Jacobian $\jacobian$ evaluated at $\x_*$ have negative real parts. For general values of $i(t)=i$, $p(t)=p$, the Jacobian is given by
\begin{equation}
	\jacobian = \begin{bmatrix}
		\partial_i h_1 & \partial_p h_1 \\
		\partial_i h_2 & \partial_p h_2
	\end{bmatrix} = \begin{bmatrix}
		\frac{R_0(1-p)-1}{\tau} & -\frac{R_0 i}{\tau} \\
		f_0 (1-p) & -if_0 - g_0
	\end{bmatrix}.
\end{equation}
If we insert $\x_*$ into $\jacobian$, we get
\begin{equation}
	\jacobian_* = \begin{bmatrix}
		0 & \frac{R_0 \frac{g_0}{f_0}(R_0-1)}{\tau} \\
		\frac{f_0}{R_0} & - g_0 R_0
	\end{bmatrix}.
\end{equation}
Then, we can compute the eigenvalues $\mu_{1,2}$ by setting the determinant of $\jacobian_*$ to zero, i.e.,
\begin{equation}
	\mathrm{det}\left( \jacobian \right) = \begin{vmatrix}
		0-\mu & \frac{R_0 \frac{g_0}{f_0}(R_0-1)}{\tau} \\
		\frac{f_0}{R_0} & - g_0 R_0 - \mu
	\end{vmatrix} = 0.
\end{equation}
Multiplying this out, gives a quadratic equation, namely,
\begin{equation}
	\mu^2 + (g_0R_0)\mu + \frac{f_0}{R_0} \cdot \frac{R_0 \frac{g_0}{f_0}(R_0-1)}{\tau} = 0,
\end{equation}
which can be solved to
\begin{equation}
	\mu_{1/2} = \frac{1}{2} \left( -g_0 R_0 \pm \sqrt{ (g_0R_0)^2 - 4 \cdot \frac{g_0(R_0-1)}{\tau} } \right).
\end{equation}
First, we consider $\mu_1$:
\begin{alignat}{2}
	\mu_1 &< 0 \\
	\sqrt{ (g_0R_0)^2 - 4 \cdot \frac{g_0(R_0-1)}{\tau} }  &< g_0 R_0 \\
	(g_0R_0)^2 - 4 \cdot \frac{g_0(R_0-1)}{\tau} &< (g_0 R_0)^2 \\
	- 4 \cdot \frac{g_0(R_0-1)}{\tau} &< 0.
\end{alignat}
Because $R_0>1$, $g_0 \geq 0$, and $\tau>0$, we see that $\mu_1$ is negative for any positive $g_0$, i.e., whenever there is at least some tendency to stop adopting \ac{PB}.

For $\mu_2$, we differentiate between whether it is real- or complex-valued. If it is real-valued, we know that $\sqrt{\cdot} \leq g_0 R_0$ and thus, $\mu_2$ is negative for any $g_0,R_0>0$. If it is complex valued, i.e., the radicand is negative, we have $\mathfrak{Re}\{\mu_2\} = - \frac{1}{2} g_0 R_0$. Thus, we can see that the equilibrium is indeed stable for all reasonable parameter values if there is at least some cost associated with \ac{PB}.

\section{Coupled Behavior-Disease Model: Sensitivity Analysis}
\begin{figure}
    \centering
    \includegraphics[width=0.5\textwidth]{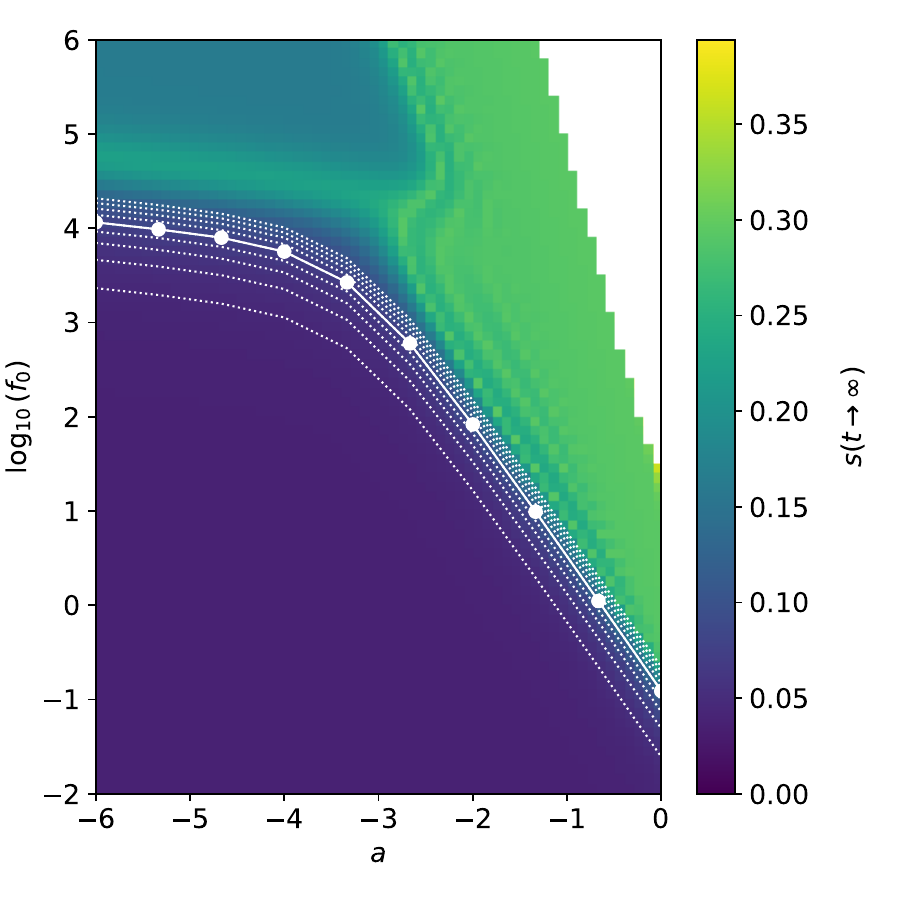}
    \caption{Sensitivity analysis for Fig.~\ref{fig:coupled_model_analysis_f0-a-plane_sfinal} with varied values of $\beta$.}
    \label{fig:coupled_model_analysis_sensitvity_beta}
\end{figure}
\begin{figure}
    \centering
    \includegraphics[width=0.5\textwidth]{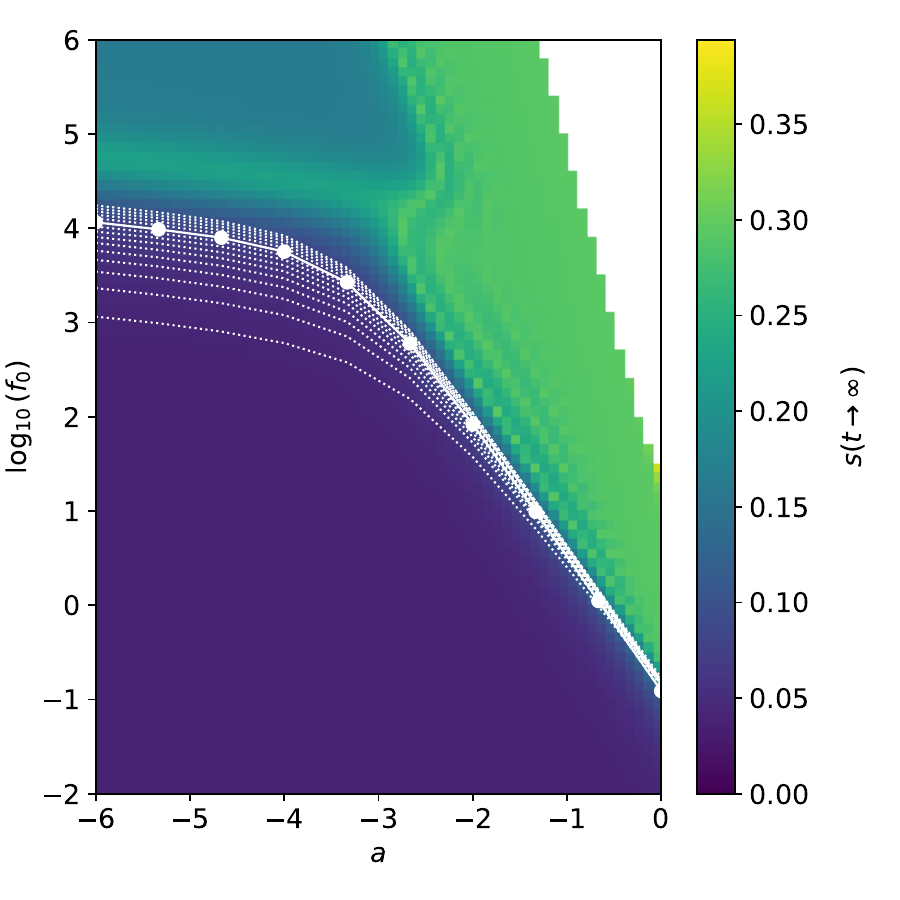}
    \caption{Sensitivity analysis for Fig.~\ref{fig:coupled_model_analysis_f0-a-plane_sfinal} with varied values of $\istar$.}
    \label{fig:coupled_model_analysis_sensitvity_istar}
\end{figure}
\begin{figure}
    \centering
    \includegraphics[width=0.5\textwidth]{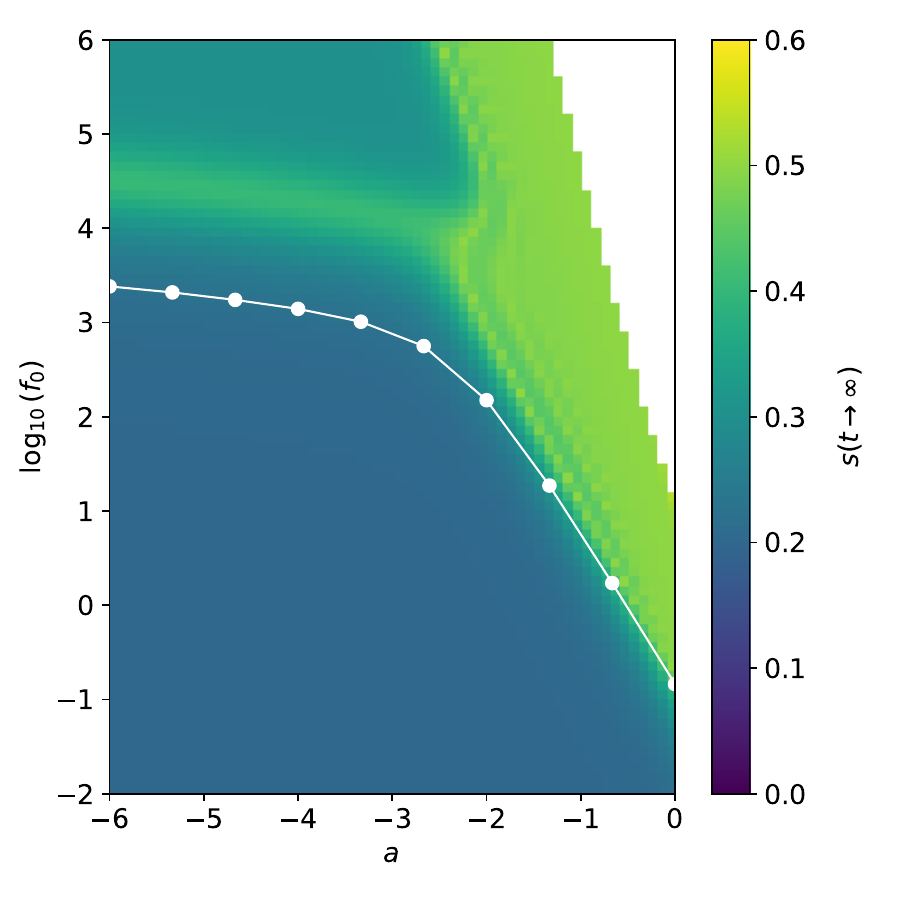}
    \caption{Sensitivity analysis for Fig.~\ref{fig:coupled_model_analysis_f0-a-plane_sfinal} with $R_0=2$.}
    \label{fig:coupled_model_analysis_sensitvity_R0}
\end{figure}
\begin{figure}
    \centering
    \includegraphics[width=0.5\textwidth]{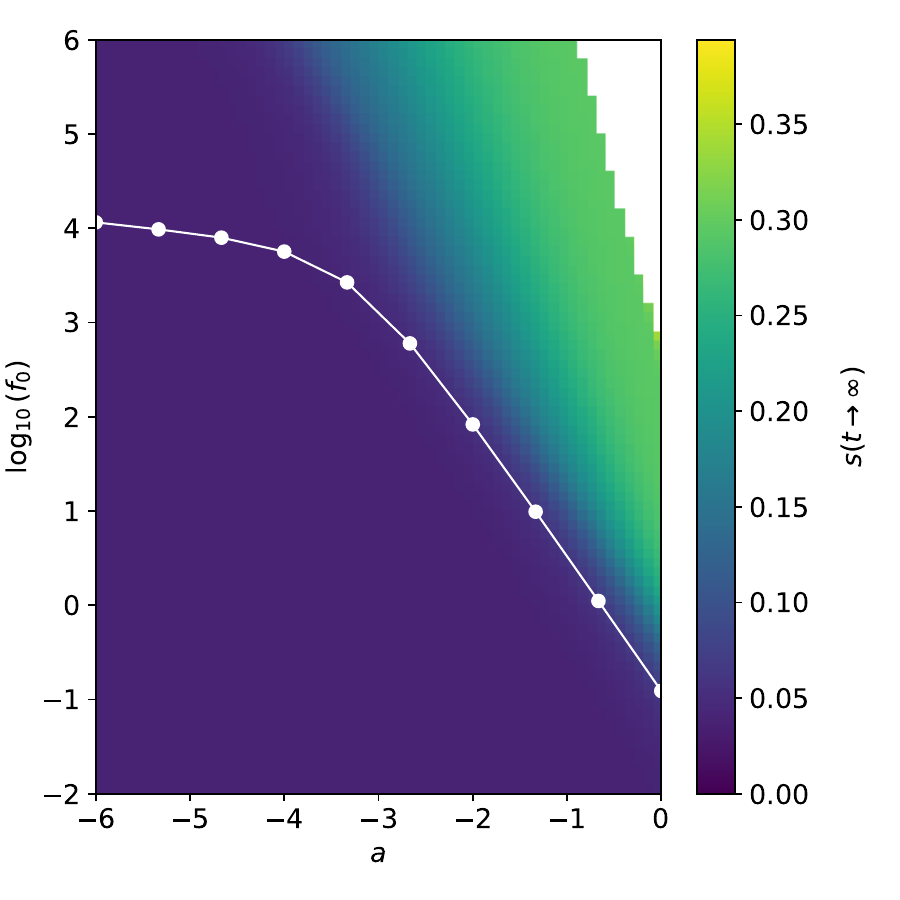}
    \caption{Sensitivity analysis for Fig.~\ref{fig:coupled_model_analysis_f0-a-plane_sfinal} with $\logten{g_0}=-1.0$.}
    \label{fig:coupled_model_analysis_sensitvity_g0}
\end{figure}

To verify that the critical curve we identified in the $a$-$f_0$-plane is not arbitrary, we provide additional sensitivity analyses here. In Fig.~\ref{fig:coupled_model_analysis_sensitvity_beta}, we take the same parameters as for the plot in the main text, but vary $\beta \in \{0.1, 0.2, \dots, 0.9\}$. In Fig.~\ref{fig:coupled_model_analysis_sensitvity_istar}, we take the same parameters as for the plot in the main text, but vary $\istar \in \{0.05, 0.1, 0.15, \dots, 0.75\}$. We show the curve obtained for the standard values as a reference and the other values by dotted white lines. 

In Fig.~\ref{fig:coupled_model_analysis_sensitvity_R0}, we perform the analysis for the same values as for the main plot, but set $R_0$ to $2$. Because fewer individuals will get infected even in absence of any \ac{PB}, we also reduce $\istar$ to $0.1$ to obtain the critical curve. 

Finally, in Fig.~\ref{fig:coupled_model_analysis_sensitvity_g0}, we perform the same analysis as in the main plot, but for $\logten{g_0}=-1.0$. In this scenario, our method to obtain the critical curve over-estimates the ability of \ac{PB} in the first wave to have a lasting effect if $a$ is small. This is because the combination of a fast awareness decay and high cost of \ac{PB} enables the emergence of a second wave that is essentially not mitigated by \ac{PB}. Because our method to obtain the critical curve essentially considers only the first wave, these effects cannot be captured.
\end{document}